\documentclass[12pt, savetrees]{llncs}

\usepackage{amsmath, amssymb, latexsym, geometry}
\usepackage[all]{xy}
\usepackage{color}

\newcommand{\nc}[1]{\newcommand{#1}}

\nc{\Cite}[1]{\textcolor{red}{\textbf [#1]}}

\nc{\ed}{

\end{document}
}

\usepackage[all]{xy}\xyoption{all}
\long\def\forget#1\forgotten{}

\nc{\ip}[2]{#1^{\, #2}}
\nc{\spst}{\supseteq}
\nc{\Span}{\op{span}}
\nc{\M}{\op{M}}
\nc{\beq}{\begin{eqnarray*}}
\nc{\eeq}{\end{eqnarray*}}
\nc{\la}{\langle}
\nc{\ra}{\rangle}
\nc{\NF}{\op{NF}}
\nc{\my}[1]{\R{\sf (#1)}}
\nc{\SSS}{\op{SSS}}\nc{\USS}{\op{USS}}
\nc{\sub}{\subseteq}
\nc{\summitinf}{\overline{\inf}}\nc{\summitsup}{\underline{\sup}}
\nc{\set}[2]{\{#1\,:\,#2\}}
\nc{\BN}{{\mathbf{B}_N}}
\nc{\op}[1]{\operatorname{#1}}
\nc{\bbZ}{\mathbb{Z}}
\nc{\bbF}{\mathbb{F}}
\nc{\GL}{\op{GL}}
\nc{\Boo}{\mathbf{B}} 
\nc{\ph}[1]{\phantom{#1}}
\definecolor{mygreen}{rgb}{0.00,0.59,0.00}
\nc{\G}[1]{{\textcolor{mygreen}{#1}}} 
\nc{\R}[1]{{\textcolor[rgb]{0.98,0.00,0.00}{#1}}} 
\nc{\B}[1]{{\textcolor{blue}{#1}}}
\nc{\x}{\times}
\nc{\LE}{\preccurlyeq}

\nc{\inv}{^{-1}}
\nc{\bi}{\begin{itemize}}
\nc{\ei}{\end{itemize}}
\nc{\be}{\begin{enumerate}}
\nc{\ee}{\end{enumerate}}
\nc{\itm}{\item}
\nc{\Impl}{\Rightarrow}
\nc{\Iff}{\Leftrightarrow}
\nc{\ga}{\G{\boxed{\G{g}^\R{a}}}}
\nc{\gb}{\G{\boxed{\G{g}^\R{b}}}}

\newtheorem{thm}{Theorem}
\newcommand{\bthm}{\begin{thm}} \newcommand{\ethm}{\end{thm}}
\newtheorem{prop}[thm]{Proposition}
\newcommand{\bprp}{\begin{prop}} \newcommand{\eprp}{\end{prop}}
\newtheorem{fact}[thm]{Fact}
\newcommand{\bfct}{\begin{fact}} \newcommand{\efct}{\end{fact}}
\newtheorem{prob}[thm]{Problem}
\newcommand{\bprb}{\begin{prob}} \newcommand{\eprb}{\end{prob}}
\newtheorem{lem}[thm]{Lemma}
\newcommand{\blem}{\begin{lem}} \newcommand{\elem}{\end{lem}}
\newcommand{\bclm}{\begin{claim}} \newcommand{\eclm}{\end{claim}}
\newtheorem{cor}[thm]{Corollary}
\newcommand{\bcor}{\begin{cor}} \newcommand{\ecor}{\end{cor}}
\newtheorem{conj}[thm]{Conjecture}
\newcommand{\bcnj}{\begin{conj}} \newcommand{\ecnj}{\end{conj}}
\newtheorem{defn}[thm]{Definition}
\newcommand{\bdfn}{\begin{defn}} \newcommand{\edfn}{\end{defn}}
\newtheorem{cnv}[thm]{Convention}
\newcommand{\bcnv}{\begin{cnv}} \newcommand{\ecnv}{\end{cnv}}
\newtheorem{ntn}[thm]{Notation}
\newcommand{\bntn}{\begin{ntn}} \newcommand{\entn}{\end{ntn}}
\newtheorem{alg}[thm]{Algorithm}
\newcommand{\balg}{\begin{alg}} \newcommand{\ealg}{\end{alg}}
\newtheorem{rem}[thm]{Remark}
\newcommand{\brem}{\begin{rem}} \newcommand{\erem}{\end{rem}}
\newtheorem{exam}[thm]{Example}
\newcommand{\bexm}{\begin{exam}} \newcommand{\eexm}{\end{exam}}
\newcommand{\bpf}{\begin{proof}} \newcommand{\epf}{\end{proof}}

\title{Polynomial-time solutions of computational problems 
in noncommutative-algebraic cryptography}

\titlerunning{Polytime cryptanalysis of noncommutative KEPs}

\author{Boaz Tsaban}
\authorrunning{Boaz Tsaban}
\institute{Department of Mathematics, Bar-Ilan University, Ram\-at Gan 52900, Israel\\
\email{tsaban@math.biu.ac.il}\\
\texttt{http://www.cs.biu.ac.il/\~{}tsaban}
}

\forget
\keywords{noncommutative-algebraic cryptography, group theory-based cryptography, braid-based cryptography,
Commutator key exchange, Centralizer key exchange, Braid Diffie--Hellman key exchange,
linear cryptanalysis, invertibility lemma, Schwartz--Zippel lemma, linear centralizer method, braid infinimum reduction,
algebraic cryptanalysis.
}
\forgotten

\begin{document}

\pagestyle{headings}

\maketitle

\begin{abstract}
We introduce the \emph{linear centralizer method},
and use it to devise a provable polynomial time solution
of the Commutator Key Exchange Problem, the computational
problem on which, in the passive adversary model,
the security of the Anshel--Anshel--Goldfeld 1999 \emph{Commutator} key exchange protocol is based.
We also apply this method to solve, in polynomial time,
the computational problem underlying the \emph{Centralizer} key exchange protocol,
introduced by Shpilrain and Ushakov in 2006.

This is the first provable polynomial time cryptanalysis of the Commutator key exchange protocol,
hitherto the most important key exchange protocol in the realm of noncommutative-algebraic cryptography,
and the first cryptanalysis (of any kind) of the Centralizer key exchange protocol.
Unlike earlier cryptanalyses of the Commutator key exchange protocol,
our cryptanalyses cannot be foiled by changing the distributions used in the protocol.

\end{abstract}

\section{Introduction}

Since Diffie and Hellman's 1976 key exchange protocol, few alternative proposals for  key exchange protocols (KEPs) resisted cryptanalysis.
This, together with the (presently, theoretical) issue that the Diffie--Hellman and other classic KEPs
can be broken in polynomial time by quantum computers, is a strong motivation
for searching for substantially different KEPs.
Lattice-based KEPs \cite{Lat} seem to be a viable potential alternative.
All classic KEPs as well as the Lattice-based ones are based on commutative algebraic structures.

In 1999, Anshel, Anshel, and Goldfeld \cite{AAG} (cf.\ \cite{AAFG}) introduced the
\emph{Commutator KEP}, a general method for constructing KEPs based on
\emph{noncommutative} algebraic structures.
Around the same time, Ko, Lee, Cheon, Han, Kang, and Park \cite{KL+00}
introduced the \emph{Braid Diffie--Hellman KEP}, another general method
achieving the same goal.
The security of both KEPs is based on variations of the \emph{Conjugacy Search Problem}:
Given conjugate elements $g,h$ in a noncommutative group, find $x$ in that group such that
$x\inv gx=h$.
Both papers \cite{AAG} and \cite{KL+00} proposed to use the \emph{braid group} $\BN$,
a finitely presented, infinite noncommutative group parameterized by a natural number $N$,
as the platform group.

The introduction of the Commutator KEP and the Braid Diffie--Hellman KEP
was followed by a stream of heuristic attacks
(e.g., \cite{HT02}, \cite{Hughes02}, \cite{LL02}, \cite{HofSte03}, \cite{BGG05}, \cite{MSU05}, \cite{Geb05}, \cite{Geb06}, \cite{Ka06}, \cite{Maffre06}, \cite{MSU06}, \cite{MU07}, \cite{MU08}),\footnote{Surveys of some of the heuristic attacks are provided in
Dehornoy \cite{DehSurv} and Garber \cite{GarSurv}.}
demonstrating that these protocols, \emph{when using the two most simple distributions} on the braid group $\BN$,
are insecure.
Consequently, a program was set forth, by several independent research groups,
to find efficiently sampleable distributions on the braid group that,
when used with the above-mentioned protocols, foil all heuristic attacks
(e.g., \cite{Maffre06}, \cite{KLT07}, \cite{GMMU08}, \cite{AnKo12}).
The abstract of \cite{GMMU08} concludes: ``Proper choice \dots{} produces a key exchange scheme which is
resistant to all known attacks''.
Moreover, a very practical distribution is announced in \cite{GagtaTalk}, which foils the strongest known
methods for solving the Conjugacy Search Problem in $\BN$.

Most of the mentioned heuristic attacks address the Commutator KEP,
and not the Braid Diffie--Hellman KEP.
The reason is that in 2003, Cheon and Jun published an expected polynomial time
cryptanalysis of the Braid Diffie--Hellman KEP, using a novel representation theoretic method \cite{CJ03}.
In their paper, Cheon and Jun stress that their cryptanalysis \emph{does not apply
to the Commutator KEP} and that an extra ingredient is needed.
Thus far, no expected polynomial time attack was found on the
Commutator KEP, whose success does not depend on the distributions used in the protocols.

The main result of the present paper is a Las Vegas, provable expected polynomial time
solution of the Commutator Key Exchange Problem (also referred to as the \emph{Anshel--Anshel--Goldfeld Problem} \cite[\S 15.1.2]{MSUbook}), the computational problem
underlying the Commutator KEP. This forms a cryptanalysis of
the Commutator KEP \cite{AAG}, in the passive adversary model, that succeeds regardless of the distributions used
to generate the keys.

The \emph{linear centralizer} method, developed for our solution of the
Commutator Key Exchange Problem, is applicable to additional computational problems and KEPs in the context of group theory-based
cryptography. We present an application of these
methods to the \emph{Centralizer KEP}, introduced by Shpilrain and Ushakov in 2006 \cite{ShUsh06}, to obtain
an expected polynomial time attack. This is the first cryptanalysis, of any kind, of the Centralizer KEP.

We stress that the cryptanalyses presented here, like the Cheon--Jun cryptanalysis, while of
expected polynomial time, are impractical for standard values of $N$
(e.g., $N=100$). These results are of theoretic nature.
Ignoring logarithmic factors, the complexity of our cryptanalyses is about $N^{17}$, times a cubic polynomial in the other
relevant parameters.
Incidentally, though, these cryptanalyses establish the first provable practical attacks in the case where the index $N$ of the braid group $\BN$ is small,
e.g., when $N=8$.

\medskip

The paper is organized as follows.
Section 2 introduces the Commutator KEP and the braid group.
In Section 3, we eliminate a technical complexity theoretic obstacle.
Section 4 applies a method of Cheon and Jun to reduce our problem to matrix groups over finite fields.
Section 5 is the main ingredient of our cryptanalysis, presenting the new method and
cryptanalyzing the Commutator KEP in  matrix groups. This section is independent of the other sections and
readers without prior knowledge of the braid group may wish to read it first.
Section 6 fills a gap in our proof, by applying the Schwartz--Zippel Lemma to obtain a lower bound on
the probability that certain random matrices are invertible.
Section 7 is a cryptanalysis of the Centralizer KEP, using the methods introduced in the earlier sections.
The Braid Diffie--Hellman KEP is introduced in Section 8, where we survey the Cheon--Jun
cryptanalysis and explain why it does not apply to the Commutator KEP or to the Centralizer KEP.
We also describe applications of the new methods to a generalized version of the Braid Diffie--Hellman KEP and to Stickel's KEP.
Some additional discussion is provided in Section 9.

\section{The Commutator KEP and the braid group $\BN$}

We will use, throughout, the following basic notation.

\bntn
For a noncommutative group $G$ and group elements $g,x\in G$,
$g^x = x\inv g x$, the conjugate of $g$ by $x$.
\entn
Useful identities involving this notation, that are easy to verify, include $g^{xy}=(g^x)^y$, and $g^c=g$ for every
\emph{central} element $c\in G$, that is, such that $ch=hc$ for all $h\in G$.

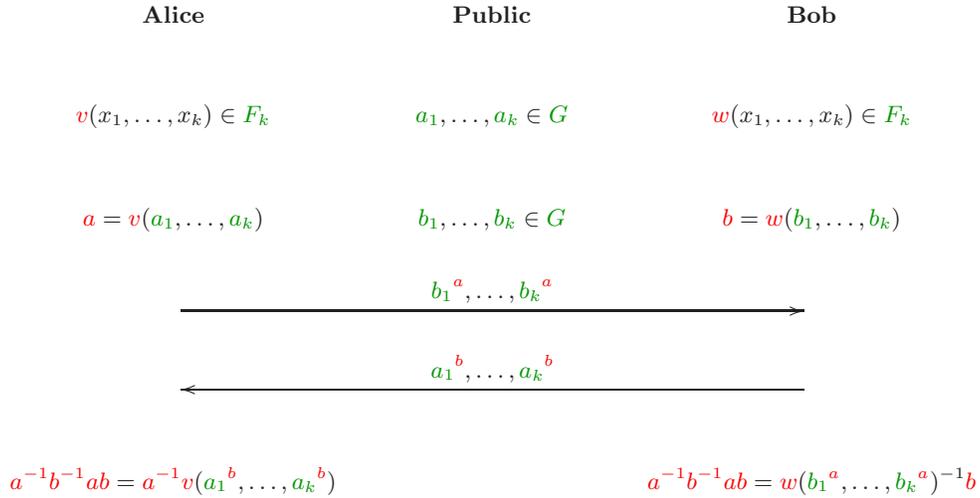
\begin{figure}[!h]
\begin{center}
$\xymatrix{
\textbf{Alice} & \textbf{Public} & \textbf{Bob}\\
\R{v}(x_1,\dots,x_k)\in \G{F_k} & \G{a_1},\dots,\G{a_k}\in \G{G} & \R{w}(x_1,\dots,x_k)\in \G{F_k}\\
\R{a}=\R{v}(\G{a_1},\dots,\G{a_k}) & \G{b_1},\dots,\G{b_k}\in \G{G} & \R{b}=\R{w}(\G{b_1},\dots,\G{b_k})\\
\ar[rr]^{\displaystyle \G{b_1}^\R{a},\dots,\G{b_k}^\R{a}} & &\\
& & \ar[ll]_{\displaystyle \G{a_1}^\R{b},\dots,\G{a_k}^\R{b}}\\
\R{a\inv b\inv ab}= \R{a\inv}\R{v}(\G{a_1}^\R{b},\dots,\G{a_k}^\R{b}) & & \R{a\inv b\inv ab}= \R{w}(\G{b_1}^\R{a},\dots,\G{b_k}^\R{a})\inv\R{b}
}$
\end{center}
\caption{The Commutator KEP}\label{fig:CKE}
\end{figure}

The \emph{Commutator KEP} \cite{AAG} is described succinctly in Figure \ref{fig:CKE}.\footnote{In our
diagrams, green letters indicate publicly known elements, and red
ones indicate secret elements, known only to the secret holders. Results of computations involving
elements of both colors may be either publicly known, or secret, depending on the context.
The colors are not necessary to follow the diagrams.} In some detail:
\be
\item A noncommutative group $G$ and elements $a_1,\dots,a_k,b_1,\dots,b_k\in G$ are publicly given.\footnote{By adding elements,
if needed, we assume that the number of $a_i$'s is equal to the number of $b_i$'s.}
\item Alice and Bob choose free group words in the variables $x_1,\dots,x_k$, $v(x_1,\dots,x_k)$ and $w(x_1,\dots,x_k)$,
respectively.\footnote{A free group word in the variables $x_1,\dots,x_k$ is a product of the form
$x_{i_1}^{\;\epsilon_1}x_{i_2}^{\;\epsilon_2}\cdots x_{i_m}^{\;\epsilon_m}$, with
$i_1,\dots,i_m\in\{1,\dots,k\}$ and $\epsilon_1,\dots,\epsilon_m\in\{1,-1\}$, and with no subproduct of the form $x_ix_i\inv$ or $x_i\inv x_i$.}
\item Alice substitutes $a_1,\dots,a_k$ for $x_1,\dots,x_k$, to obtain a secret element $a=v(a_1,\dots,a_k)\in G$.
Similarly, Bob computes $b=w(b_1,\dots,b_k)\in G$.
\item Alice sends the conjugated elements ${b_1}^a,\dots,{b_k}^a$ to Bob, and Bob sends
${a_1}^b,\dots,{a_k}^b$ to Alice.
\item The shared key is the \emph{commutator} $a\inv b\inv ab$.
\ee
As conjugation is a group isomorphism, we have that
$$v({a_1}^b,\dots,{a_k}^b)=v({a_1},\dots,{a_k})^b=a^b=b\inv a b.$$
Thus, Alice can compute the shared key $a\inv b\inv ab$ as $a\inv v({a_1}^b,\dots,{a_k}^b)$, using her secret $a, v(x_1,\dots,\allowbreak x_k)$ and
the public elements ${a_1}^b,\dots,{a_k}^b$.
Similarly, Bob computes $a\inv b\inv ab$ as $w({b_1}^a,\dots,{b_k}^a)\inv b$.

For the platform group $G$, it is proposed in \cite{AAG} to use the \emph{braid group} $\BN$, a group parameterized by a natural number $N$.
Elements of $\BN$ are called \emph{braids}, for they may be identified with braids on $N$ strands.
Braid group multiplication is motivated geometrically, but the details will play no role in the present paper.
The interested reader will find detailed information on $\BN$ in almost each of the papers in the bibliography
and, in particular, in the survey \cite{BirmanBrendle}, but prior knowledge is not necessary:
we quote here the information needed for the present paper.

Let $S_N$ be  the symmetric group of permutations on $N$ symbols.
For our purposes, the braid group $\BN$ is a group of elements of the form
$$(i,\mathbf{p}),$$
where $i$ is an integer, and $\mathbf{p}$ is a finite (possibly, empty) sequence of elements of $S_N$, that is,
$\mathbf{p}=(p_1,\dots,p_\ell)$ for some $\ell\ge 0$ and $p_1,\dots,p_\ell\in S_N$.
The sequence $\mathbf{p}=(p_1,\dots,p_\ell)$ is requested to be \emph{left weighted} (a property whose definition
will not be used here), and $p_1$ must not be the involution $p(k)=N-k+1$.\footnote{For
readers familiar with the braid group, we point out that the sequence $(i,(p_1,\dots,p_\ell))$ encodes the left normal form $\Delta^ip_1\cdots p_\ell$
of the braid, in Artin's presentation, with $\Delta$ being the fundamental, half twist braid on $N$ strands.}

For ``generic'' braids $(i,(p_1,\dots,p_\ell))\in\BN$, $i$ is negative and $|i|$ is $O(\ell)$, but this is not always the case.
Note that the bit-length of an element $(i,(p_1,\dots,p_\ell))\in\BN$ is $O(\log|i|+\ell N\log N)$.

Multiplication is defined on $\BN$ by an algorithm of complexity $O(\ell^2 N\log N+\log|i|)$.
Inversion is of linear complexity.
Explicit implementations are provided, for example, in \cite{CK+01}.

For a passive adversary to extract the shared key of the Commutator KEP out of the public information,
it suffices to solve the following problem, also referred to as the \emph{Anshel--Anshel--Goldfeld Problem} \cite[\S 15.1.2]{MSUbook}.

\bprb[Commutator KEP Problem]\label{prb:CKE}
Let $a_1,\dots,a_k,b_1,\dots,b_k\in \BN$, each of the form $(i,\mathbf{p})$ with $\mathbf{p}$ of length $\le\ell$.
Let $a$ be a product of at most $m$ elements of $\{a_1,\dots,a_k\}^{\pm 1}$,
and let $b$ be a product of at most $m$ elements
of $\{b_1,\dots,b_k\}^{\pm 1}$.\\
Given $a_1,\dots,a_k,b_1,\dots,b_k,a_1^b,\dots,a_k^b,b_1^a,\dots,b_k^a$, compute $a\inv b\inv ab$.
\eprb

Our solution of Problem \ref{prb:CKE} consists of several ingredients.

\section{Reducing the infimum}\label{sec:Inf}

The \emph{infimum} of a braid $b=(i,\mathbf{p})$ is the integer $\inf(b):=i$.
As the bit-length of $b$ is $O(\log|i|+\ell N\log N)$, an algorithm polynomial in $|i|$ would be at least \emph{exponential} in the bit-length.
We first remove this obstacle.

In cases where $\mathbf{p}$ is the empty sequence, we write $(i)$ instead of $(i,\mathbf{p})$. The properties of $\BN$
include, among others, the following ones.
\be
\itm[(a)] $(i)\cdot(j,\mathbf{p})=(i+j,\mathbf{p})$ for all integers $i$ and all $(j,\mathbf{p})\in\BN$.\\
In particular, $(i)=(1)^i$ for all $i$.
\itm[(b)] $(2)\cdot (i,\mathbf{p})=(i,\mathbf{p})\cdot (2)$ for all for all $(i,\mathbf{p})\in\BN$.
\ee
Thus, $(2j)$ is a central element of $\BN$ for each integer $j$.
If follows that, for each $(i,\mathbf{p})\in\BN$,
$$(i,\mathbf{p}) = (i-(i \bmod 2))\cdot (i \bmod 2,\mathbf{p}).$$
This way, every braid $b\in\BN$ decomposes to a unique product $c\tilde b$, where $c$ is of the form $(2j)$ (and thus \emph{central}), and
$\inf(\tilde b)\in\{0,1\}$.

Consider the public information in Figure \ref{fig:CKE}.
For each $j=1,\dots,k$, decompose as above
\beq
a_j & = & c_j\tilde a_j,\\
b_j & = & d_j\tilde b_j,
\eeq
with $c_j,d_j$ central and $\inf(\tilde a_j),\inf(\tilde b_j)\in\{0,1\}$ for all $j=1,\dots,k$.
Let
\beq
\tilde a & = & v(\tilde a_1,\dots,\tilde a_k);\\
\tilde b & = & w(\tilde b_1,\dots,\tilde b_k);\\
c & = & v(c_1,\dots,c_k);\\
d & = & w(d_1,\dots,\tilde d_k).
\eeq
As the elements $c_j,d_j$ are central, we have that
$$\tilde a =  v(c_1\inv a_1,\dots,c_k\inv a_k) = v(c_1\inv,\dots,c_k\inv)\cdot v(a_1,\dots, a_k) = c\inv a.$$
Similarly, $\tilde b=d\inv b$.
As $c$ and $d$ are central,
$$\ip{a_j}{b} = (c_j\tilde a_j)^b = c_j \ip{\tilde a_j}{b} = c_j\ip{\tilde a_j}{d\tilde b} = c_j\ip{\tilde a_j}{\tilde b}$$
for all $j=1,\dots,k$.
Thus, $\tilde a_j^{\tilde b}$ can be computed for all $j$.
Similarly, $\ip{\tilde b_j}{\tilde a}$ can be computed.
Now,
$$a\inv b\inv ab = (c\tilde a)\inv (d\tilde b)\inv (c\tilde a) (d\tilde b) = \tilde a\inv c\inv  \tilde b\inv d\inv c\tilde a d\tilde b
=  \tilde a\inv \tilde b\inv \tilde a \tilde b.$$
This shows that the Commutator KEP Problem 
is reducible, in linear time,
to the same problem using $\tilde a_1,\dots,\tilde a_k,\tilde b_1,\dots,\tilde b_k$
instead of $a_1,\dots,a_k,b_1,\dots,b_k$.
Thus, we may assume that
$$\inf(a_1),\dots,\inf(a_k),\inf(b_1),\dots,\inf(b_k)\in\{0,1\}$$
to start with. Assume that henceforth.

For a braid $x=(i,\mathbf{p})$, let $\ell(\mathbf{p})$ be the number of permutations in the sequence $\mathbf{p}$.
For integers $i,s$, let
$$[i,s]=\set{x\in\BN}{i\le\inf(x)\le\inf(x)+\ell(x)\le s}.$$
We use the following basic facts about $\BN$:
\be
\item If $x_1\in [i_1,s_1]$ and $x_2\in [i_2,s_2]$, then $x_1x_2\in[i_1+i_2,s_1+s_2]$.
\item If $x\in [i,s]$, then $x\inv\in [-s,-i]$.
\ee
Thus, for each $x\in\{a_1,\dots,a_k,b_1,\dots,b_k\}^{\pm 1}$,
$x^{\pm 1}\in [-\ell-1,\ell +1]$, and therefore, in the notation of our problem,
$a,b\in [-m(\ell +1),m(\ell +1)]$. Thus,
$$a\inv b\inv ab\in  [-4m(\ell +1),4m(\ell +1)].$$

\bcor\label{cor:bound}
In the Commutator KEP Problem, 
$a\inv b\inv ab\in  [-4m(\ell +1),4m(\ell +1)]$.
\ecor

\section{Reducing to a matrix group over a finite field}\label{sec:rep}

In this section, we apply methods of Cheon and Jun \cite{CJ03} in our setting. 

Let $n$ be a natural number. As usual, we denote the algebra of all $n\x n$ matrices over a field $\bbF$ by $\M_n(\bbF)$, and
the group of invertible elements of this algebra by $\GL_n(\bbF)$.
A \emph{matrix group} is a subgroup of $\GL_n(\bbF)$.
A \emph{faithful representation} of a group $G$ in $\GL_n(\bbF)$ is a group isomorphism from $G$ onto a matrix group
$H\le \GL_n(\bbF)$. A group is \emph{linear} if it has a faithful representation.

Bigelow and, independently, Krammer, established in their breakthrough papers  \cite{Bigelow01}, \cite{Krammer02}
that the braid group $\BN$ is linear, by proving that the so-called \emph{Lawrence--Krammer representation}
$$\op{LK}\colon \BN\longrightarrow \GL_{\binom{N}{2}}(\bbZ[t^{\pm 1},\frac{1}{2}]),$$
whose dimension is
$$n:=\binom{N}{2},$$
is injective.\footnote{Bigelow proved this theorem for the coefficient ring $\bbZ[t^{\pm 1},q^{\pm 1}]$ with
two variables. Krammer proved, in addition, that one may replace $q$ by any real number from the interval $(0,1)$.}
The Lawrence--Krammer representation of a braid
can be computed in polynomial time.\footnote{When the infimum $i$ is polynomial in the other parameters, which we
proved in Section \ref{sec:Inf} that we may assume. Alternatively, by computing the representation of $(i)$ separately,
using properties of the Lawrence--Krammer representation.}
It is proved implicitly in \cite{Krammer02}, and explicitly in \cite{CJ03},
that this representation is also invertible in (similar) polynomial time.
The following result follows from Corollary 1 of \cite{CJ03}.

\bthm[Cheon--Jun \cite{CJ03}]\label{thm:CJ1}
Let $x\in [i,s]$ in $\BN$. Let $M\ge\max(|i|,|s|)$. Then:
\be
\itm The degrees of $t$ in $\op{LK}(x)\in \GL_n(\bbZ[t^{\pm 1},\frac{1}{2}])$ are in $\{-M,-M+1,\dots,M\}$.
\itm The rational coefficients $\frac{c}{2^d}$ in $\op{LK}(x)$ ($c$ integer, $d$ nonnegative integer) satisfy: $|c|\le 2^{N^2M}, |d|\le 2NM$.
\ee
\ethm

In the notation of Theorem \ref{thm:CJ1}, Theorem 2 in Cheon--Jun \cite{CJ03} implies that inversion of $\op{LK}(x)$ is of
order $N^6\log M$ multiplications of entries. Ignoring logarithmic factors and thus assuming that each entry multiplication costs
$NM\cdot N^2M=N^3M^2$, this accumulates to $N^8M^2$.
We will invert the function $\op{LK}$ as part of our cryptanalysis below. However, the
complexity of the other steps in our cryptanalysis (in particular, the
linear centralizer step---Section \ref{sec:LC}) dominate the complexity of inverting $\op{LK}$.

Let us return to  the Commutator KEP Problem \ref{prb:CKE}.
By Corollary \ref{cor:bound},
$$K:=a\inv b\inv ab\in  [-4m(\ell +1),4m(\ell +1)].$$
Let $M=4m(\ell+1)$.
By Theorem \ref{thm:CJ1}, we have that
$$(2^{2NM}t^M)\cdot\op{LK}(K)\in\GL_n(\bbZ[t]),$$
the absolute values of the coefficients in this matrix are bounded by $2^{N^2(M+1)}$,
and the maximal degree of $t$ in this matrix is bounded by $2M$.

Let $p$ be a prime slightly greater than $2^{N^2M+2NM}$, and $f(t)$ be an irreducible polynomial over $\bbZ_p$, of degree $d$ slightly larger than $2M$.
Then
$$(2^{2NM}t^M)\cdot\op{LK}(K)=(2^{2NM}t^M)\cdot\op{LK}(K) \bmod (p, f(t))
\in \GL_{n}(\bbZ[t]/\langle p, f(t)\rangle),$$
under the natural identification of $\{-(p-1)/2,\dots,(p-1)/2\}$ with $\{0,\dots,p-1\}$.

Let $\bbF=\bbZ[t]/\langle p, f(t)\rangle = \bbZ[t^{\pm 1},\frac{1}{2}]/\langle p, f(t)\rangle$.
$\bbF$ is a finite field of cardinality $p^d$, where $d$ is the degree of $f(t)$.
It follows that the complexity of field operations in $\bbF$ is, up to logarithmic factors, of order
$$d^2\log p = O(M^3 N^2) = O(m^3\ell^3N^2).$$
Thus, the key $K$ can be recovered as follows:
\be
\itm Apply the composed function $\op{LK}(x) \bmod (p, f(t))$ to the input of the Commutator KEP Problem,
to obtain a version of this problem in $\GL_n(\bbF)$.
\itm Solve the problem there, to obtain $\op{LK}(K) \bmod (p, f(t))$.
\itm Compute $(2^{2NM}t^M)\cdot\op{LK}(K) \bmod (p, f(t)) = (2^{2NM}t^M)\cdot\op{LK}(K)$.\footnote{The equality here
is over the integers.}
\item Divide by $(2^{2NM}t^M)$ to obtain $\op{LK}(K)$.
\item Compute $K$ using the Cheon--Jun inversion algorithm.
\ee
It remains to devise a polynomial time solution of the Commutator KEP Problem in arbitrary groups of matrices.

\section{Linear centralizers}\label{sec:LC}

In this section, we solve the Commutator KEP Problem in matrix groups.
We first state the problem in a general form. As usual, for a group $G$ and
elements $g_1,\dots,g_k\in G$, $\langle g_1,\dots,g_k \rangle$ denotes the subgroup of $G$ generated by $g_1,\dots,g_k$.
Throughout, we assume that the given groups are represented in an efficient way.

\bprb[Commutator KEP Problem]\label{prb:CKEG}
Let $G$ be a group. Let $a_1,\dots,a_k,b_1,\dots,b_k\in G$.
Let $a\in\langle a_1,\dots,a_k\rangle, b\in \langle b_1,\dots,b_k\rangle$.\\
Given $a_1,\dots,a_k,b_1,\dots,b_k,a_1^b,\dots,a_k^b,b_1^a,\dots,b_k^a$, compute $a\inv b\inv ab$.
\eprb

We recall a classic definition.

\bdfn
Let $S\sub\M_n(\bbF)$ be a set.
The \emph{centralizer} of $S$ (in $\M_n(\bbF)$) is the set
$$C(S)=\set{c\in \M_n(\bbF)}{cs = sc\mbox{ for all }s\in S}.$$
For $a_1,\dots,a_k\in \M_n(\bbF)$, $C(\{a_1,\dots,a_k\})$ is also denoted as $C(a_1,\dots,a_k)$.
\edfn

Basic properties of $C(S)$, that are easy to verify, include:
\be
\item $C(S)$ is a vector subspace (indeed, a matrix subalgebra) of $\M_n(\bbF)$.
\item $C(C(S))\spst S$.
\item $C(S)=C(\Span S)$.
\item If $S\sub\GL_n(\bbF)$, then $C(S)=C(\langle S\rangle)$, where $\langle S\rangle$ is the subgroup of $\GL_n(\bbF)$ generated by $S$.
\ee

A key observation is the following one:
Let $V$ be a vector subspace of $\M_n(\bbF)$,
and $G\le\GL_n(\bbF)$ be a matrix group such that $V\cap G$ is nonempty. It may be computationally
infeasible to find an element in $V\cap G$.
However, it is easy to compute a basis for $V\cap U$ for any vector subspace $U$ of $\M_n(\bbF)$. In particular,
this is true for $U=C(C(G))$, that contains $G$. In certain cases, as the ones below,
a ``random'' element in $V\cap C(C(G))$ is as good as one in $V\cap G$.

Algorithm \ref{alg:CKEP} below addresses the Commutator KEP Problem in a matrix group $G\le\GL_n(\bbF)$.
The analysis of this algorithm is based on the forthcoming Lemma \ref{invmx},
which shows that one can efficiently find an invertible matrix in a vector space of matrices containing at
least one invertible matrix.
To this end, we assume that  $|\bbF|/n\ge c>1$ for some constant $c$. In the above section, $|\bbF|/n$ is at least exponential.
Fix a finite set $S\sub\bbF$ of cardinality greater than $cn$ (the larger the better),
that can be sampled efficiently.
In the most important case, where $\bbF$ is a finite field, take $S=\bbF$.
By \emph{random element} of a vector subspace $V$ of $\M_n(\bbF)$,
with a prescribed basis $\{v_1,\dots,v_d\}$, we mean a linear combination
$$\alpha_1 v_1+\cdots+\alpha_kv_k$$
with $\alpha_1,\dots,\alpha_k\in S$ uniform, independently distributed.

It is natural to split the Commutator KEP Problem and the algorithm
for solving it into an offline (preprocessing) phase and an online phase.

\balg\label{alg:CKEP}\mbox{}\\
\emph{Offline phase:}
\be
\item \emph{Input:} $b_1,\dots,b_k\in G$.
\item \emph{Execution:}
\be
\item Compute a basis $S=\{s_1,\dots,s_d\}$ for $C(b_1,\dots,b_k)$,
by solving  the following homogeneous system of linear equations
in the $n^2$ entries of the unknown matrix $x$:
$$
\begin{array}{rcl}
b_1\cdot x & = & x\cdot b_1\\
& \vdots\\
b_k\cdot x & = & x\cdot b_k.
\end{array}
$$
\item Compute a basis for $C(S)=C(C(b_1,\dots,b_k))$,
by solving  the following homogeneous system of linear equations
in the $n^2$ entries of the unknown matrix $x$:
$$
\begin{array}{rcl}
s_1\cdot x & = & x\cdot s_1\\
& \vdots\\
s_d\cdot x & = & x\cdot s_d.
\end{array}
$$
\ee
\item \emph{Output:} A basis for $C(C(b_1,\dots,b_k))$.
\ee
\emph{Online phase:}
\be
\item \emph{Input:} $a_1,\dots,a_k,b_1,\dots,b_k,a_1^b,\dots,a_k^b,b_1^a,\dots,b_k^a\in G$, where
$a\in\langle a_1,\dots,a_k\rangle, b\in \langle b_1,\dots,b_k\rangle$ are unknown.
\item \emph{Execution:}
\be
\item Solve the following homogeneous system of linear equations
in the $n^2$ entries of the unknown matrix $x$:
$$
\begin{array}{rcl}
b_1\cdot x & = & x\cdot{b_1}^{a}\\
& \vdots\\
b_k\cdot x & = & x\cdot{b_k}^{a}.
\end{array}
$$
\item Fix a basis for the solution space, and pick random solutions $x$ until $x$ is invertible.
\item Solve the following homogeneous system of linear equations
in the $n^2$ entries of the unknown matrix $y$:
$$
\begin{array}{rcl}
a_1\cdot y & = & y\cdot{a_1}^{b}\\
& \vdots\\
a_k\cdot y & = & y\cdot{a_k}^{b},
\end{array}
$$
subject to the \emph{linear constraint} that $y\in C(C(b_1,\dots,b_k))$.
\item Fix a basis for the solution space, and  pick random solutions $y$ until $y$ is invertible.
\item \emph{Output:} $x\inv y\inv xy$.
\ee
\ee
\ealg

Let $\omega$ be the matrix multiplication constant, that is, the minimal such that matrix multiplication is $O(n^{\omega+o(1)})$.
For our applications, one may take $\omega=\log_2 7\approx 2.81$.
As usual, \emph{Las Vegas algorithm} means an algorithm that always outputs the correct answer in finite time.
For the proof of the following theorem, note that if $g^x=g^y$, then $g^{xy\inv}=g$, or in other words,
$xy\inv\in C(g)$. Finally, note that it does not make much sense to consider the case where $k>n^2$, in which the matrices
become linearly dependent and thus redundant.

\bthm\label{thm:CKE}
Assume that $|\bbF|/n\ge c>1$ for some constant $c$, and $k\le n^2$.
Algorithm \ref{alg:CKEP} is a Las Vegas algorithm for the Commutator KEP Problem, 
with running time, in units of field operations:
\be
\item \emph{Offline phase:} $O(n^{2\omega+2})$.
\item \emph{Online phase:} $O(kn^{2\omega})$.
\ee
\ethm
\bpf
We use the notation of Algorithm  \ref{alg:CKEP}.
First, assume that the algorithm terminates.
We prove that its output is $a\inv b\inv ab$.
$$x\inv y\inv xy = x\inv y\inv (x a\inv) ay.$$
The equations 2(a) in the online phase of Algorithm \ref{alg:CKEP} assert that ${b_i}^x={b_i}^a$ for all $i=1,\dots,k$.
Thus, $xa\inv\in C(b_1,\dots,b_k)$. As $y\in C(C(b_1,\dots,b_k))$, $y$ commutes with $xa\inv$,
and therefore so does $y\inv$.
Thus,
$$x\inv y\inv (x a\inv) ay = x\inv (xa\inv)y\inv ay = a\inv y\inv ay = a\inv a^y.$$
By the equations 2(c) in the online phase of Algorithm \ref{alg:CKEP}, ${a_i}^y={a_i}^b$ for all $i=1,\dots,k$.
As $a\in\langle a_1,\dots,a_k\rangle$, we have that $a^y=a^b$. Indeed, let
$a=a_{i_1}^{\epsilon_1}\cdots a_{i_m}^{\epsilon_m}$. As conjugation is an isomorphism,
$$a^y = (a_{i_1}^{\epsilon_1})^y\cdots (a_{i_m}^{\epsilon_m})^y =
({a_{i_1}^y})^{\epsilon_1}\cdots ({a_{i_m}^y})^{\epsilon_m} =
({a_{i_1}^b})^{\epsilon_1}\cdots ({a_{i_m}^b})^{\epsilon_m} =
(a_{i_1}^{\epsilon_1})^b\cdots (a_{i_m}^{\epsilon_m})^b =
a^b.$$
Thus,
$$a\inv a^y=a\inv a^b = a\inv b\inv ab.$$
Thus, the algorithm returns the correct answer when it terminates. It remains to analyze the running
time of the algorithm, which we do step-by-step.

Offline phase, Step  2(a):
These are $kn^2$ equations in $n^2$ variables, and thus the running time is $O(k(n^2)^\omega)=O(kn^{2\omega})$.

Offline phase, Step 2(b):
As $C(b_1,\dots,b_k)$ is a vector subspace of $\M_n(\bbF)$, its dimension $d$ is at most $n^2$.
Thus, the running time of this step is $O(n^2\cdot n^{2\omega}) = O(n^{2\omega+2})$.

Online phase, Step  2(a):
The running time is $O(kn^{2\omega})$, as in Step 2(a) of the offline phase.

Online phase, Step  2(b):
There is an invertible solution to the equations 2(a), namely: $a$.
Thus, by the Invertibility Lemma (Lemma \ref{invmx} below),
the probability that a random solution is \emph{not} invertible may be assumed arbitrarily close to
$n/|\bbF|\le 1/c<1$.
Thus, the expected number of random elements picked until an invertible one is found is constant.
To generate one random element, one takes a linear combination of a basis of the solution space. If $m$ is the dimension,
then $m\le n^2$ and the linear combination takes $mn^2\le n^4$ operations. Checking invertibility is faster.
The total expected running time of this step is, therefore, $O(n^4)$, and $n^4\le n^{2\omega}$.

Online phase, Step  2(c):
Let $\{s_1,\dots,s_m\}$ be the basis computed in the offline phase. Then $m\le n^2$.
In the present step, one sets $y=t_1s_1+\cdots+t_ms_m$, with $t_1,\dots,t_m$ variables, and obtains
$kn^2$ equations in the $m\le n^2$ variables $t_1,\dots,t_m$.
The complexity is $O(\frac{kn^2}{m}m^\omega)$, and
$\frac{kn^2}{m}m^\omega=kn^2\cdot m^{\omega-1}\le kn^{2\omega}$.

Online phase, Step  2(d):
Using the same arguments as in Step 2(b), the running time of this step is $O(n^{2\omega})$.
\epf

\section{Finding an invertible solution when there is one}

The results in the previous section assume that we are able to find,
efficiently, an invertible matrix in any subspace of $\M_n(\bbF)$ containing an invertible element.
This is taken care of by the following Lemma.

\blem[Invertibility Lemma]\label{invmx}
Let $a_1,\dots,a_m\in\M_n(\bbF)$ be such that
$$\op{span}\{a_1,\dots,a_m\}\cap \GL_n(\bbF)\neq \emptyset.$$
Let $S$ be a finite subset of $\bbF$.
If $\alpha_1,\dots,\alpha_m$ are chosen uniformly and independently from $S$, then the probability
that $\alpha_1a_1+\cdots+\alpha_ma_m$ is invertible is at least $1-\frac{n}{|S|}$.
\elem
\bpf
Let
$$f(t_1,\dots,t_m)=\det(t_1a_1+\cdots+t_ma_m)\in \bbF[t_1,\dots,t_m],$$
where $t_1,\dots,t_m$ are scalar variables.
This is a determinant of a matrix whose coefficients are linear in the variables.
By the definition of determinant as a sum of products of $n$ elements,
$f$ is a polynomial of degree $n$.
As $\op{span}\{a_1,\dots,a_m\}\cap \GL_n(\bbF)\neq \emptyset$,
$f$ is nonzero.

The proof is completed by applying the Schwartz--Zippel Lemma (Lemma \ref{SchZ} below).
\epf

For the reader's convenience, we include a proof for the following classic lemma.

\blem[Schwartz--Zippel]\label{SchZ}
Let $f(t_1,\dots,t_m)\in \bbF[t_1,\dots,t_m]$ be a nonzero multivariate polynomial of degree $n$.
Let $S$ be a finite subset of $\bbF$.
If $\alpha_1,\dots,\alpha_m$ are chosen uniformly and independently from $S$,
then the probability that $f(\alpha_1,\dots,\alpha_m)\neq 0$ is at least $1-\frac{n}{|S|}$.
\elem
\bpf
We prove the lemma by induction on $m$.

If $m=1$, then $f$ is a univariate polynomial of degree $n$, and thus has at most $n$ roots.

For the inductive step, assume that $m>1$ and write
$$f(t_1,\dots,t_m)=f_0(t_2,\dots,t_m)+f_1(t_2,\dots,t_m)t_1+f_2(t_2,\dots,t_m)t_1^2+\cdots+f_k(t_2,\dots,t_m)t_1^k,$$
with $k\le n$ maximal such that $f_k(t_2,\dots,t_m)$ is nonzero.
The degree of $f_k(t_2,\dots,t_k)$ is at most $m-k$.
For each choice of $\alpha_2,\dots,\alpha_m\in\bbF$ with $f_k(\alpha_2,\dots,\alpha_m)\neq 0$,
$f(t_1,\alpha_2,\dots,\alpha_m)$ is a univariate polynomial of degree $k$ in the variable $t_1$.
By the induction hypothesis (for $m=1$), for random $\alpha_1\in S$, $f(\alpha_1,\alpha_2,\dots,\alpha_m)$ is nonzero with probability at least $1-k/|S|$.
By the induction hypothesis,
\nc{\pr}[1]{\op{Pr}\bigl[#1\bigr]}
\beq
\lefteqn{\pr{f(\alpha_1,\dots,\alpha_m)\neq 0} \ge}\\
 & \ge & \pr{f_k(\alpha_2,\dots,\alpha_m)\neq 0}\cdot \pr{f(\alpha_1,\dots,\alpha_m)\neq 0\mid f_k(\alpha_2,\dots,\alpha_m)\neq 0}\ge\\
& \ge & \left(1-\frac{n-k}{|S|}\right)\left(1-\frac{k}{|S|}\right)\ge  1-\frac{n}{|S|}.
\eeq
\epf

\section{Application to the Centralizer KEP}\label{sec:Centralizer}

\bdfn
For a group $G$ and an element $g\in G$, the \emph{centralizer of $g$ in $G$} is the set
$$C_G(g):=\set{h\in G}{gh=hg}.$$
\edfn

\begin{figure}[!h]
\begin{center}
$\xymatrix@R=15pt{
\textbf{Alice} & \textbf{Public} & \textbf{Bob}\\
\R{a_1}\in \G{G} & \G{g}\in\G{G} & \R{b_2}\in\G{G}\\
\ar[rr]^{\displaystyle \G{g_1},\dots,\G{g_k}\in C_\G{G}(\R{a_1})} & &\\
& & \ar[ll]_{\displaystyle \G{h_1},\dots,\G{h_k}\in C_\G{G}(\R{b_2})}\\
\R{a_2}\in \langle \G{h_1},\dots,\G{h_k} \rangle & & \R{b_1}\in \langle \G{g_1},\dots,\G{g_k} \rangle\\
\ar[rr]^{\displaystyle \R{a_1}\G{g}\R{a_2}} & &\\
& & \ar[ll]_{\displaystyle \R{b_1}\G{g}\R{b_2}}\\
\R{K}= \R{a_1}\R{b_1}\G{g}\R{b_2}\R{a_2} & & \R{K} = \R{b_1}\R{a_1}\G{g}\R{a_2}\R{b_2}
}$
\end{center}
\caption{The Centralizer KEP}\label{fig:CentKEP}
\end{figure}
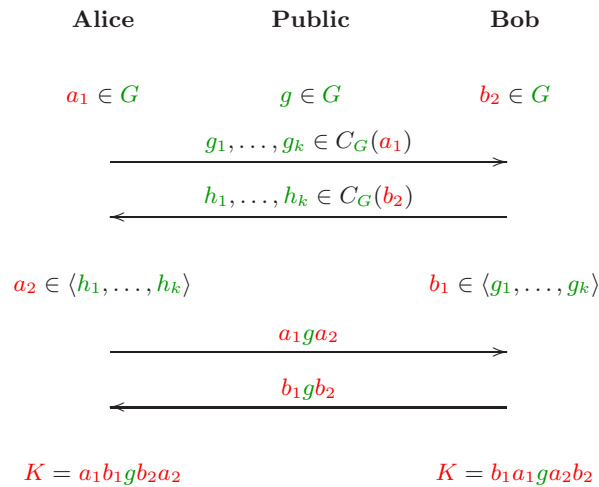

The \emph{Centralizer KEP}, introduced by Shpilrain and Ushakov in 2006 \cite{ShUsh06}, is described in Figure \ref{fig:CentKEP}.
In this protocol, $a_1$ commutes with $b_1$ and $a_2$ commutes with $b_2$.
Consequently, the keys computed by Alice and Bob are identical, and equal to $a_1b_1ga_2 b_2$.

As in the Commutator KEP, it is proposed in  \cite{ShUsh06} to use the braid group $\BN$ as the platform group $G$.
The group elements are chosen in a special way, so as to foil attacks attempted at earlier braid group based KEPs.
We apply the methods developed in the previous sections to obtain an expected polynomial time cryptanalysis of this KEP.
We omit some details, that are similar to those in the earlier sections.

\bprb[Centralizer KEP Problem]\label{prb:CentKEP}
Assume that $g,a_1,b_2\in \BN$, $g_1,\dots,g_k \in \break C_\BN(a_1)$, $h_1,\dots,h_k\in C_\BN(b_2)$,
each of the form $(i,\mathbf{p})$ with $\mathbf{p}$ of length $\le\ell$.
Let $a_2$ be a product of at most $m$ elements
of $\{h_1,\dots,h_k\}^{\pm 1}$,
and let $b_1$ be a product of at most $m$ elements of $\{g_1,\dots,g_k\}^{\pm 1}$.\\
Given $g,g_1,\dots,g_k,h_1,\dots,h_k, a_1ga_2,b_1gb_2$, compute $a_1b_1ga_2b_2$.
\eprb

\subsection{Solving the Centralizer KEP Problem in matrix groups}
For a group $G$, $Z(G)=C_G(G)$ is the set of all central elements of $G$.
Consider the variation of the Centralizer KEP Problem \ref{prb:CentKEP}, where the group is
$G\le\GL_n(\bbF)$ instead of $\BN$.
The following variation of this problem is formally harder than this variation.\footnote{Since
$Z(G)\sub C_G(g)$ for every $g\in G$,
the mentioned problems are, under mild technical hypotheses perhaps, equivalent.
We will not use this feature.}

\bprb\label{prb:MxCentKEP}
Let $G\le\GL_n(\bbF)$. Assume that $g,a_1,b_2\in G$, $g_1,\dots,g_k\in C_G(a_1)$, $h_1,\dots,h_k\in C_G(b_2)$,
$a_2\in \langle \{h_1,\dots,h_k\}\cup Z(G)\rangle$, and $b_1\in\langle \{g_1,\dots,g_k\}\cup Z(G)\rangle$.\\
Given $g,g_1,\dots,g_k, h_1,\allowbreak\dots, h_k, \allowbreak a_1ga_2,b_1gb_2$, compute $a_1b_1ga_2b_2$.
\eprb

Following is an algorithm for solving Problem \ref{prb:MxCentKEP}.
As before, for $S\sub\M_n(\bbF)$, $C(S)$ (without subscript)
is the centralizer of $S$ \emph{in the matrix algebra} $\M_n(\bbF)$.

\balg\label{alg:CentKEP}\mbox{}
\be
\item \emph{Input:} $g,g_1,\dots,g_k,h_1,\dots,h_k, a_1ga_2,b_1gb_2\in G$.
\item \emph{Execution:}
\be
\item Compute bases for the subspaces $C(g_1,\dots,g_k)$, $C(C(h_1,\dots,h_k))$ of $\M_n(\bbF)$.
\item Solve
$$x\cdot g = a_1ga_2\cdot y$$
subject to the linear constraints $x\in C(g_1,\dots,g_k), y\in C(C(h_1,\dots,h_k))$.
\item Take random linear combinations of the basis of the solution space to obtain solutions $(x,y)$,
until $y$ is invertible.
\ee
\item \emph{Output:} $x\cdot b_1gb_2\cdot y\inv$.
\ee
\ealg

\bthm
Let $G\le\GL_n(\bbF)$. Assume that $|\bbF|/n\ge c>1$ for some constant $c$, and $k\le n^2$.
Algorithm \ref{alg:CentKEP} is a Las Vegas algorithm for Problem \ref{prb:MxCentKEP},
with running time, in units of field operations, $O(n^{2\omega+2})$.
\ethm
\bpf
The proof is similar to that of Theorem \ref{thm:CKE}.

First, assume that the algorithm terminates.
We prove that its output is $a_1b_1 g a_2 b_2$.
As $x\in C(g_1,\dots,g_k)$ and $b_1\in\langle g_1,\dots,g_k\rangle$, $x$ commutes with $b_1$.
As $b_2$ commutes with $h_1,\dots,h_k$,  $b_2\in C(h_1,\dots,h_k)$.
As $y\in C(C(h_1,\dots,h_k))$, $y$ commutes with $b_2$, and therefore so does $y\inv$.
Thus,
$$xb_1gb_2y\inv = b_1xgy\inv b_2.$$
As $xg=a_1ga_2y$, $xgy\inv = a_1ga_2$. Thus,
$$b_1xgy\inv b_2 = b_1a_1ga_2b_2 = a_1b_1ga_2b_2.$$

The analysis of the running time of the algorithm is essentially identical to the analysis in Theorem \ref{thm:CKE}.
In Step 2(c), let
$$H=\set{(x,y)\in  C(g_1,\dots,g_k)\x C(C(h_1,\dots,h_k))}{x\cdot g = a_1ga_2\cdot y}$$
be the solution space, and let $(x_1,y_1),\dots,(x_d,y_d)$ be a basis for $H$.
As $H$ is a subspace of $\M_n(\bbF)\x \M_n(\bbF)$, $d\le 2n^2$.
Let $H_2=\set{y}{(x,y)\in H}$, the projection of $H$ on the second coordinate.
Then
$$H_2=\Span\{y_1,\dots,y_d\}.$$
$(a_1,a_2\inv)\in H$, and thus $a_2\inv\in H_2$. In particular, there is an invertible element in $H_2$.
By the Invertibility Lemma (Lemma \ref{invmx}), a random linear combination of $y_1,\dots,y_d$ is invertible
with probability at least $1/c$.
The total expected running time of this step is, therefore, $O(n^4)$, and $n^4\le n^{2\omega}$.
\epf

\subsection{Solving the Centralizer KEP Problem in the braid group}\label{subsec:Inf}

We now address Problem \ref{prb:CentKEP}. We begin by reducing to the case where our braids have a
restricted form.

In Section \ref{sec:Inf}, we explained how each $x\in\BN$ can be decomposed (in linear time) as $x=c\tilde x$ with $c$ central and $\inf(x)\in\{0,1\}$.

We may assume that
$$\inf(g)\in\{0,1\}.$$
Indeed, assume that we have an algorithm solving the problem when $\inf(g)\in\{0,1\}$.
Write $g=c\tilde g$ with $c$ central and $\inf(g)\in\{0,1\}$.
Compute
\beq
c\inv a_1ga_2 & = & a_1c\inv ga_2 = a_1\tilde ga_2;\\
c\inv b_1gb_2 & = & b_1c\inv gb_2 = b_1\tilde gb_2.
\eeq
Apply the given algorithm to $\tilde g, g_1,\dots,g_k,h_1,\dots,h_k, a_1\tilde ga_2,b_1\tilde gb_2$,
to obtain $a_1b_1\tilde ga_2b_2$. Multiply by $c$ to obtain  $a_1b_1 ga_2b_2$.

We may, in addition, assume that
$$\inf(g_1),\dots,\inf(g_k),\inf(h_1),\dots,\inf(h_k)\in\{0,1\},$$
since when we apply Algorithm \ref{alg:CentKEP} in the image
of our group in a matrix group, we have in Problem \ref{prb:MxCentKEP} that
\beq
\langle \{h_1,\dots,h_k\}\cup Z(G)\rangle & = & \langle \{\tilde h_1,\dots,\tilde h_k\}\cup Z(G)\rangle;\\
\langle \{g_1,\dots,g_k\}\cup Z(G)\rangle & = & \langle \{\tilde g_1,\dots,\tilde g_k\}\cup Z(G)\rangle.
\eeq
As in Section \ref{sec:Inf}, it follows that
$$a_2,b_1\in[-m(\ell+1),m(\ell+1)].$$
Let $u=a_1ga_2$ and $v=b_1gb_2$. Decompose $u=c\tilde u$ and $v=d\tilde v$ with $c,d$ central and $\inf(\tilde u),\inf(\tilde v)\in\{0,1\}$.
As $g\in [0,\ell+1]$ and $a_1\in[\inf(a_1),\inf(a_1)+\ell]$,
$$u=a_1ga_2\in [\inf(a_1),\inf(a_1)+(m+1)(\ell+1)+\ell],$$
and thus
\beq
a_1g(c\inv a_2) = \tilde u & \in & [0,(m+1)(\ell+2)];\\
c\inv a_1 = \tilde u a_2\inv g\inv & \in & [-(m+1)(\ell+1),(m+1)(2\ell+3)].
\eeq
Similarly,
$$(d\inv b_1)gb_2=\tilde v\in [0,(m+1)(\ell+2)].$$
Finally,
$$K':=a_1(d\inv b_1)gb_2(c\inv a_2) = a_1\tilde v(c\inv a_2) = (c\inv a_1) \tilde v a_2 \in [-(m+2)(\ell+1),(m+1)(4\ell+6)].$$
Let $M=(m+2)(4\ell+6)$. Continue as in Section \ref{sec:Inf}.

By Theorem \ref{thm:CJ1}, we have that
$$(2^{2NM}t^M)\cdot\op{LK}(K')\in\GL_n(\bbZ[t]),$$
the absolute values of the coefficients in this matrix are bounded by $2^{N^2(M+1)}$,
and the maximal degree of $t$ in this matrix is bounded by $2M$.
Let $p$ be a prime slightly greater than $2^{N^2M}$, and $f(t)$ be an irreducible polynomial over $\bbZ_p$, of degree $d$ slightly larger than $2M$.
Then
$$(2^{2NM}t^M)\cdot\op{LK}(K')=(2^{2NM}t^M)\cdot\op{LK}(K') \bmod (p, f(t))
\in \GL_{n}(\bbZ[t]/\langle p, f(t)\rangle),$$
under the natural identification of $\{-(p-1)/2,\dots,(p-1)/2\}$ with $\{0,\dots,p-1\}$.
Let $\bbF=\bbZ[t]/\langle p, f(t)\rangle = \bbZ[t^{\pm 1},\frac{1}{2}]/\langle p, f(t)\rangle$.
$\bbF$ is a finite field of cardinality $p^d$, where $d$ is the degree of $f(t)$.
It follows that the complexity of field operations in $\bbF$ is, up to logarithmic factors, of order
$$d^2\log p = O(M^3 N^2) = O(m^3\ell^3N^2).$$
Thus, the key $K$ can be recovered as follows:
\be
\itm Apply the composed function $\op{LK}(x) \bmod (p, f(t))$ to
$$g,g_1,\dots,g_k,h_1,\dots,h_k, \tilde u=\allowbreak a_1g(c\inv a_2),\tilde v=(d\inv b_1)gb_2,$$
to obtain an input to Problem \ref{prb:MxCentKEP}.
\itm Solve the problem there, to obtain $\op{LK}(K') \bmod (p, f(t))$.
\itm Compute $(2^{2NM}t^M)\cdot\op{LK}(K') \bmod (p, f(t)) = (2^{2NM}t^M)\cdot\op{LK}(K')$.
\item Divide by $(2^{2NM}t^M)$ to obtain $\op{LK}(K')$.
\item Compute $K'$ using the Cheon--Jun inversion algorithm.
\item Multiply by $cd$ to obtain $a_1b_1ga_2b_2$.
\ee

\section{Further applications}

\subsection{The Braid Diffie--Hellman KEP}

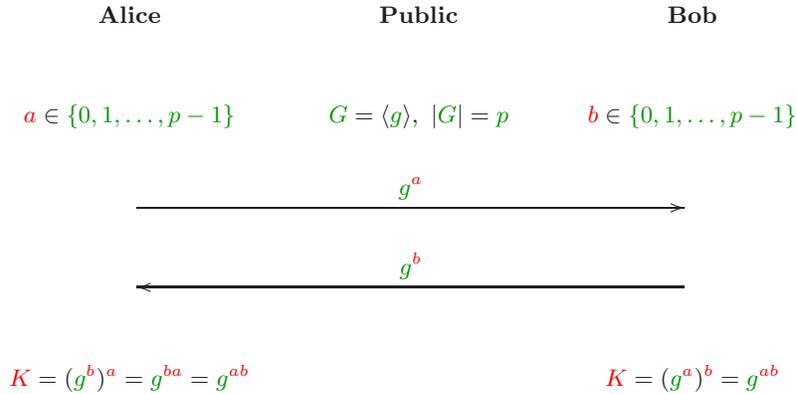
\begin{figure}[!h]
\begin{center}
$\xymatrix{
\textbf{Alice} & \textbf{Public} & \textbf{Bob}\\
\R{a}\in\G{\{0,1,\dots,p-1\}} & \G{G}=\langle \G{g}\rangle,\ |\G{G}|=\G{p} & \R{b}\in\G{\{0,1,\dots,p-1\}}\\
\ar[rr]^{\displaystyle \G{g}^\R{a}} & &\\
& & \ar[ll]_{\displaystyle \G{g}^\R{b}}\\
\R{K}= (\G{g}^\R{b})^{\R{a}} =\G{g}^{\R{ba}} =\G{g}^{\R{ab}} & & \R{K} = (\G{g}^\R{a})^{\R{b}} =\G{g}^{\R{ab}}
}$
\caption{The Diffie--Hellman KEP}\label{DHKEP}
\end{center}
\end{figure}

Figure \ref{DHKEP} illustrates the well known Diffie--Hellman KEP.
Here, $G$ is a cyclic group of prime order,
generated by a group element $g$,
and exponentiation denotes ordinary exponentiation.

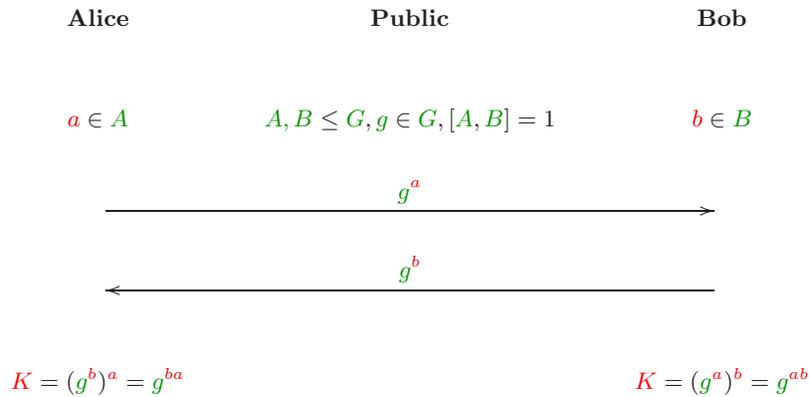
\begin{figure}[!h]
\begin{center}
$\xymatrix{
\textbf{Alice} & \textbf{Public} & \textbf{Bob}\\
\R{a}\in \G{A} & \G{A,B}\le\G{G}, \G{g}\in\G{G}, [\G{A},\G{B}]=1 & \R{b}\in \G{B}\\
\ar[rr]^{\displaystyle \G{g}^\R{a}} & &\\
& & \ar[ll]_{\displaystyle \G{g}^\R{b}}\\
\R{K}= (\G{g}^\R{b})^{\R{a}} =\G{g}^{\R{ba}} & & \R{K} = (\G{g}^\R{a})^{\R{b}} =\G{g}^{\R{ab}}
}$
\end{center}
\caption{The Braid Diffie--Hellman KEP}\label{BDHKEP}
\end{figure}

Interpreting exponentiation in noncommutative groups as conjugation leads to the
Ko--Lee--Cheon--Han--Kang--Park \emph{Braid Diffie--Hellman KEP} \cite{KL+00}.
For subsets $A,B$ of a group $G$, $[A,B]=1$
means that $a$ and $b$ commute, $ab=ba$, for all $a\in A,b\in B$.
The Braid Diffie--Hellman KEP is illustrated in Figure \ref{BDHKEP}.
Since, in the Braid Diffie--Hellman KEP, the subgroups $A$ and $B$ of $G$ commute element-wise,
the keys computed by Alice and Bob are identical.
It is proposed in \cite{KL+00} to use Artin's braid group $\BN$ as the platform group $G$
for the Braid Diffie--Hellman KEP, hence the term \emph{Braid} in the name of this KEP.

In the passive adversary model, the security of the Braid Diffie--Hellman KEP for a platform group $G$
(Figure \ref{BDHKEP}) is captured by the following problem.

\bprb[Diffie--Hellman Conjugacy Problem]
Let $A$ and $B$ be subgroups of $G$ with $[A,B]=1$, and let $g\in G$ be given.
Given a pair $(g^a, g^b)$ where $a\in A$ and $b\in B$, find $g^{ab}$.
\eprb

The Cheon--Jun attack on the  Braid Diffie--Hellman KEP \cite{CJ03} forms a solution to the Diffie--Hellman Conjugacy Problem in the case where $G$ is the  braid group $\BN$.
Their solution can be described, roughly, as follows. Using the methods described in Section \ref{sec:rep}, the problem is
reduced to the case where $G\le\GL_n(\bbF)$, a matrix group over a finite field. Since we are dealing with solutions that are
supposed to work for all problem instances, \emph{this problem is not harder than that where $G=\GL_n(\bbF)$}, the
group of all invertible matrices in $\M_n(\bbF)$. However, the latter problem is easy:
Assume that $B=\la b_1,\dots,b_k\ra\le G$. Solve the system
\beq
x g^a & = & gx\\
x b_1 & = & b_1x\\
& \vdots\\
xb_k & = & b_kx
\eeq
of $(k+1)n^2$ linear equations in the $n^2$ entries of the unknown matrix $x$. There
is an invertible solution to this system, namely, $a$.
Now, any invertible solution $\tilde a$ of this system can be used to compute $g^{ab}$:
By the first equation,
$$g^{\tilde a}=\tilde a\inv (g\tilde a) = \tilde a\inv (\tilde a g^a) = g^a.$$
By the remaining equations, $\tilde a$ commutes with the generators of $B$, and consequently with all elements of $B$.
Thus, we can compute
$$(g^b)^{\tilde a}=g^{b\tilde a} = g^{\tilde a b} = (g^{\tilde a})^b = (g^a)^b=g^{ab}.$$
This essentially establishes that the Diffie--Hellman Conjugacy Problem in this scenario can be solved in time $O(kn^{2\omega})$.

\subsubsection{Comparison with our approach.}
The reason why the above-mentioned approach of Cheon and Jun is not applicable, as is, to the Commutator KEP
or to the Centralizer KEP is that, in either case, there is no prescribed set of generators with which it
suffices that the solution commutes:
In the Commutator KEP (Figure \ref{fig:CKE}) it is not clear that $a$ has to commute with anything.
In the Centralizer KEP (Figure \ref{fig:CentKEP}), we need $a_2$  to commute with $b_2$, but $b_2$ is secret.
The main ingredient in our solution, in both cases,
is the replacement of membership in a subgroup with membership in the double centralizer
(in the full matrix algebra) of that subgroup, and the observation that the latter is efficiently computable.
In other words, \emph{instead of adding equations that guarantee that the solution commutes with prescribed elements,
we enlarge the set of solutions by moving to the double centralizer}, and prove the increase in the set of solutions is not
too large.

The secondary ingredients of our approach also have something to contribute to the Cheon--Jun attack:
First, in \cite{CJ03} the dependence of the complexity on the infimum $i$ is exponential (in the bit-length of $i$).
This can be eliminated using the infimum reduction methods of Sections \ref{sec:Inf} and \ref{subsec:Inf}.
Second, the fact that the solution to the above-mentioned system of equations is invertible with overwhelming
probability is not proved in \cite{CJ03}.\footnote{An argument involving Zariski density is provided in \cite{CJ03},
but this seems to be a heuristic argument; not one intended to be a rigorous proof.}
This gap may be filled using the Invertibility Lemma (Lemma \ref{invmx}).
Third, our approach may be used to push most of the work to the offline phase.

\bthm\label{DHCPthm}
Assume that $|\bbF|/n\ge c>1$.
The Diffie--Hellman Conjugacy Problem for a matrix group $G\le \GL_n(\bbF)$ and $B=\la b_1,\dots,b_k\ra$ is solvable in
$O(kn^{2\omega})$ offline time and $O(n^{2\omega})$ online  Las Vegas time.
More precisely, the running time of the online phase is $O(n^{2\omega})$, plus $O(n^\omega)$ Las Vegas time.
\ethm
\bpf
Offline phase: Compute a basis for the centralizer $C(B)$ in the matrix algebra $\M_n(\bbF)$,
a solution space of a system of $kn^2$ linear equations in the $n^2$ entries of the variable matrix $x$.
Since $C(B)$ is a subspace of the vector space $\M_n(\bbF)$, its dimension $d$ is at most $n^2$.
Let $c_1,\dots,c_d$ be a basis for $C(B)$.

Online phase: Given $g^a$, solve $xg^a = gx$ subject to $x\in C(B)$,
a linear system of $n^2$ equations in $d$ scalar variables.
Let $H$ be the solution space.
Let $h_1,\dots,h_{\tilde d}$ be a basis for $H$. Then $\tilde d\le d$.

There is an invertible element in $H$, namely: $a$.
By the Invertibility Lemma (Lemma \ref{invmx}),
if $t_1,\dots,t_{\tilde d}$ are chosen uniformly and independently from a large subset of $\bbF$,
then the matrix $\tilde a = t_1 h_1+\cdots +t_{\tilde d}h_{\tilde d}$ is invertible with probability at least $1/c$.
Having found such invertible $\tilde a$, compute
$$g^{\tilde a}=\tilde a\inv (g\tilde a) = \tilde a\inv (\tilde a g^a) = g^a.$$
The running time of the online phase is $O(n^{2\omega})$, plus $O(n^\omega)$ Las Vegas time
for the expected constant number of $n\x n$ matrix inversions.
\epf

\brem
In the complexity of the offline phase in Theorem \ref{DHCPthm}, $k$ can be taken to be the minimum among the number of
generators of $A$ and the number of generators of $B$, by exchanging the roles of $A$ and $B$.
\erem

\subsection{Double Coset KEPs}

\begin{figure}[!h]
\begin{center}
$\xymatrix{
\textbf{Alice} & \textbf{Public} & \textbf{Bob}\\
\R{a_1}\in \G{A_1}, \R{a_2}\in \G{A_2} & \G{A_1},\G{A_2},\G{B_1},\G{B_2}\le\G{G}, \G{g}\in\G{G}, [\G{A_i},\G{B_i}]=1 & \R{b_1}\in \G{B_1}, \R{b_2}\in \G{B_2}\\
\ar[rr]^{\displaystyle \R{a_1}\G{g}\R{a_2}} & &\\
& & \ar[ll]_{\displaystyle \R{b_1}\G{g}\R{b_2}}\\
\R{K}= \R{a_1}\R{b_1}\G{g}\R{b_2}\R{a_2} & & \R{K} = \R{b_1}\R{a_1}\G{g}\R{a_2}\R{b_2}
}$
\caption{The Double Coset KEP}\label{DCKEP}
\end{center}
\end{figure}
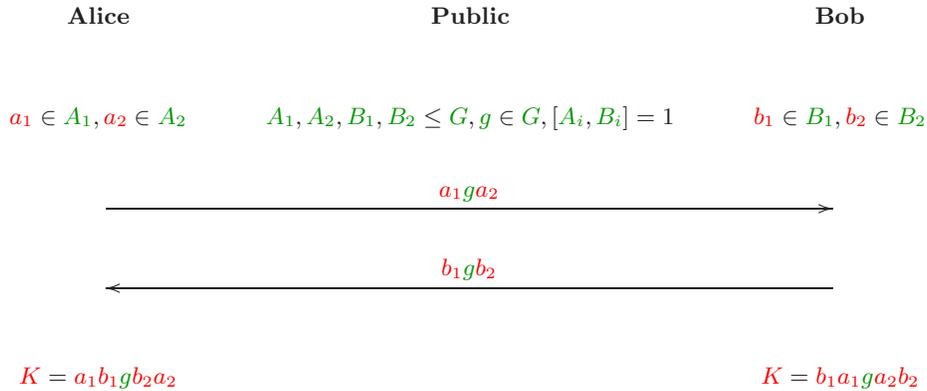

In 2001, Cha, Ko, Lee, Han and Cheon \cite{CK+01} proposed a variation of the Braid Diffie--Hellman KEP (Figure \ref{BDHKEP}).
For this variation, Cheon and Jun \cite{CJ03} described a convincing variation of their attack.
Another variation of this protocol was proposed in 2005, by Shpilrain and Ushakov \cite{ShpUshThomp}. Both variations,
as well as the Braid Diffie--Hellman KEP, are special cases of the protocol illustrated in Figure \ref{DCKEP}.

The methods of Theorem \ref{DHCPthm} extend to the Double Coset KEP.
Here too, the restriction to matrix groups is with no loss
of generality, and we obtain an expected polynomial time solution of the underlying problem in the braid group $\BN$.

\bthm\label{DCKEPthm}
Assume that $|\bbF|/n\ge c>1$.
Let $A_1,A_2,B_1=\la b_1,\dots,b_k\ra,B_2=\la b'_1,\dots,b'_l\ra\le G\le \GL_n(\bbF)$,  with $[A_1,B_1]=[A_2,B_2]=1$, and $g\in G$.
After an offline computation of complexity $O((k+l)n^{2\omega})$, one can, given
$a_1ga_2, b_1gb_2$, compute $a_1b_1ga_2b_2$ in time $O(n^{2\omega})$, plus $O(n^\omega)$ Las Vegas time.
\ethm
\bpf
Offline phase: Compute a basis for the centralizers $C(B_1), C(B_2)$
in the matrix algebra $\M_n(\bbF)$, by solving one system of $kn^2$ linear equations in $n^2$ variables, and
another system of $ln^2$ linear equations in $n^2$ variables.
Let $c_1,\dots,c_{d_1}$ be a basis for $C(B_1)$, and $c'_1,\dots,c'_{d_2}$ be a basis for $C(B_2)$. $d_1,d_2\le n^2$.

Online phase: Given $a_1ga_2$, solve $x(a_1ga_2) = gy$ subject to $x\in C(B_1), y\in C(B_2)$,
a system of $n^2$ equations in $d_1+d_2\le 2n^2$ scalar variables.
Let $H$ be the solution space,
$$H=\set{(x,y)\in C(B_1)\x C(B_2)}{x(a_1ga_2) = gy},$$
and let $(h_1,g_1),\dots,(h_d,g_d)$ be a basis for $H$. $d\le d_1+d_2\le 2n^2$.

Let $H_1=\set{x}{(xy)\in H}$ be the projection of $H$ on the first coordinate.
Then $\{h_1,\dots,h_d\}$ spans $H_1$.
There is an element $(x,y)\in H$ with $x$ (and $y$) invertible, namely: $(a_1\inv,a_2)$.
Thus, there is an invertible element in $H_1$.
By Lemma \ref{invmx}, if $t_1,\dots,t_d$ are chosen uniformly and independently from a large subset of $\bbF$,
then the matrix $x = t_1 h_1+\cdots +t_{d}h_{d}$ is invertible with probability at least $1/c$.
Let  $\tilde a_2= t_1 g_1+\cdots +t_{d}g_{d}$. Then $(x,\tilde a_2)\in H$.
Compute $\tilde a_1=x\inv$. Then
$$\tilde a_1 g\tilde a_2=x\inv (g\tilde a_2) = x\inv(xa_1ga_2) = a_1ga_2.$$
As $x\in C(B_1)$, $\tilde a_1\in C(B_1)$.
Compute
$$\tilde a_1 b_1gb_2\tilde a_2=b_1\tilde a_1 g\tilde a_2b_2= b_1 a_1 g a_2b_2=a_1b_1  g a_2b_2.$$
\epf

An interesting further application is to Stickel's KEP \cite{Stickel}.
This KEP was cryptanalyzed by Shpilrain in \cite{Sh08}, describing a heuristic cryptanalysis and
supporting it by experimental results. Stickel's KEP is a special case of the Double Coset KEP, where $G=\GL_n(\bbF)$,
$A_1=B_1=\la \{a\}\cup Z(G)\ra$, and $A_2=B_2=\la b\ra$ ($a,b$ public). By Theorem \ref{DCKEPthm},
Shpilrain's cryptanalysis can be turned into a provable Las Vegas algorithm that runs in expected polynomial time, i.e., one supported
by a rigorous mathematical proof. In particular, in this way, it is guaranteed that changing the distributions according
to which the protocol chooses the involved group elements would not defeat the mentioned polynomial time cryptanalysis.

\section{Additional comments}

Ignoring logarithmic factors, the overall complexity of the algorithms presented here
is $n^{2\omega+2}=N^{4\omega+4}$ field operations, that are of complexity $m^3\ell^3 N^2$. Thus, the complexity of our
algorithms is
$$N^{4\omega+6} m^3\ell^3,$$
ignoring logarithmic factors. While polynomial, this complexity is practical only for braid groups of small index $N$.
However, these algorithms constitute the first provable polynomial time cryptanalyses of the Commutator KEP and
of the Centralizer KEP.

The main novelty of our approach lies in the usage of linear centralizers (and double centralizers). However,
also the secondary ingredients of our analysis may be of interest. In particular, we have shown that
the Invertibility Lemma can be used to turn the Cheon--Jun cryptanalysis of the Braid Diffie--Hellman KEP \cite{CJ03}
and the Shpilrain cryptanalysis of Stickel's KEP \cite{Sh08} into provable Las Vegas algorithm that runs in expected polynomial time,
and that the infimum reduction method can be applied to the Cheon--Jun attack to eliminate the exponential dependence on the
bit-length of the infimum.

The major challenge is to reduce the degree of $N$ in the polynomial time cryptanalyses.
By Chinese Remaindering or $p$-adic lifting methods, it may be possible to reduce the complexity contributed by the field operations.
Apparently, this may reduce the power of $N$ by $1$.
It should be possible to make sure that the Invertibility Lemma is still applicable when these methods are used.
Much of the complexity comes from the Lawrence--Krammer representation having dimension quadratic in $N$.
Unfortunately, it is conjectured that there are no faithful representations of $\BN$ of smaller dimension.
A more careful analysis of the Lawrence--Krammer representation may yield finer estimates.
However, it does not seem that any of these directions would make the attacks practical for, say, $N=100$.

\begin{figure}[!h]
\begin{center}
$\xymatrix{
\textbf{Alice} & \textbf{Public} & \textbf{Bob}\\
\R{a,a_1,a_2,x_1,x_2} &
\G{\begin{matrix}
A & A_1 & A_2 & X_1 & X_2 &\\
 & | & | & | & |\\
   & Y_1 & Y_2 & B_1 & B_2 & B
\end{matrix}}
\le\G{G} &
\R{y_1,y_2,b_1,b_2,b}\\
\ar[rr]^{\displaystyle \R{\phantom{{}_1\inv\!\!\!\!\!\!}ax_1}, \R{x_1\inv a_1x_2}, \R{x_2\inv a_2}} & &\\
& & \ar[ll]_{\displaystyle \R{\phantom{{}_1\inv\!\!\!\!\!\!}b_1y_1}, \R{y_1\inv b_2y_2}, \R{y_2\inv b}}\\
\R{K}=\R{a}\R{\phantom{{}_1\inv\!\!\!\!\!\!}b_1y_1}\R{a_1}\R{y_1\inv b_2y_2}\R{a_2}\R{y_2\inv b}
& & \R{K}=\R{\phantom{{}_1\inv\!\!\!\!\!\!}ax_1}\R{b_1}\R{x_1\inv a_1x_2}\R{b_2}\R{x_2\inv a_2}\R{b}
}$
\caption{The Triple Decomposition KEP}\label{fig:TDK}
\end{center}
\end{figure}
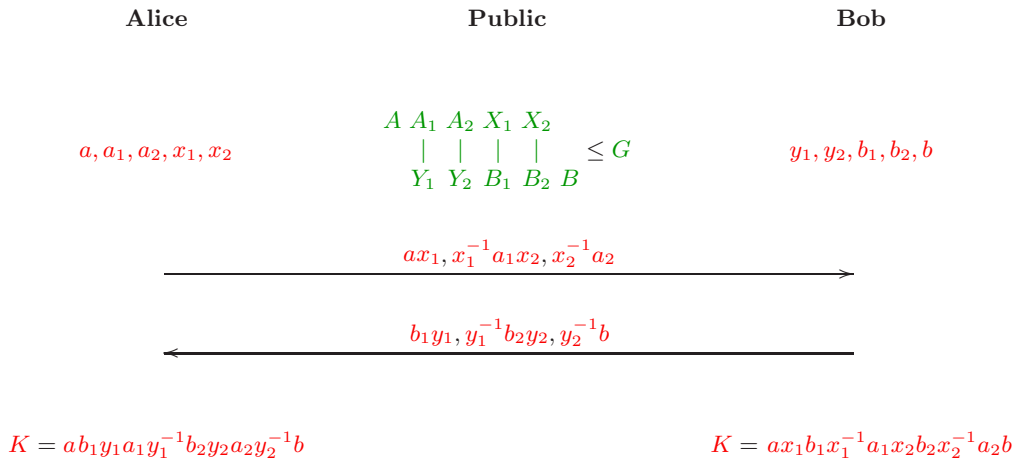

One may wonder whether, from the \emph{complexity theoretic} point of view,
this paper may be the end of braid-based cryptography.
Our belief is that this is not the case.
For example, consider Kurt's \emph{Triple Decomposition KEP} (\cite{Kurt}, \cite[4.2.5]{MSUbook}), described in Figure \ref{fig:TDK}.
In this figure, an edge between two subgroups means that these subgroups commute elementwise. This ensures that the
keys computed by Alice and Bob are both equal to $ab_1a_1b_2a_2b$.

We do not, at present, know whether the Triple Decomposition KEP
can be cryptanalyzed using the methods presented here, or whether there is a provable, efficient cryptanalysis at all.
Additional KEPs to which the present methods do not seem
to be applicable are introduced by Kalka in \cite{Ka07} and \cite{Ka12}.
There are additional types of braid-based schemes (e.g., authentication schemes),
that cannot be attacked using the methods presented here.
Some examples are reviewed in the monograph \cite{MSUbook}.

Changing the platform group in any of the studied KEPs is a very interesting option. There are efficiently implementable, infinite groups
with no faithful representations as matrix groups (e.g., the braided Thompson group).\footnote{As
for \emph{finite} nonabelian groups, we are pessimistic. For example, finite simple groups tend to be linear of small dimension,
by the classification of finite simple groups, and our method would reduce the cryptanalysis to the problem of finding an \emph{efficient}
linear representation of small dimension.
There are at present no signs that such representations must be harder to evaluate (or invert) than, say,
computing discrete logarithms in $\bbZ_p^*$. Indeed, results of Babai, Beals, and Seress seem to indicate
otherwise \cite{BBS09}.
}

\subsection*{Acknowledgments}
I worked on the Commutator KEP, from various other angles,
since I was introduced to it at the Hebrew University CS Theory seminar, by
Alex Lubotzky \cite{LubLec01}.
I thank Oleg Bogopolski for inviting me, earlier this year (2012),
to deliver a minicourse \cite{GagtaTalk} in the conference
\emph{Geometric and Combinatorial Group Theory with Applications} (D\"usseldorf, Germany, July 25--August 3, 2012). Preparing this minicourse, I discovered the linear centralizer attack.
Initially, I addressed the Centralizer KEP (Section \ref{sec:Centralizer}). When I moved to consider
the Commutator KEP, Arkadius Kalka pointed out an obstacle mentioned by Shpilrain and Ushakov,
that struck me as solvable by linear centralizers.
I am indebted to Kalka for making the right comment at the right time.

I also thank David Garber, Arkadius Kalka, and Eliav Levy, and the referees,
for comments leading to improvements in the presentation of this paper.

\ed
\begin{thebibliography}{00}

\bibitem{AnKo12} B. An, K. Ko,
\emph{A family of pseudo-Anosov braids with large conjugacy invariant sets},
ArXiv eprint arXiv:1203.2320, 2012.

\bibitem{AAG} I. Anshel, M. Anshel, D. Goldfeld, \emph{An algebraic method for public-key cryptography},
Mathematical Research Letters \textbf{6} (1999), 287--291.

\bibitem{AAFG} I. Anshel, M. Anshel, B. Fisher, D. Goldfeld,
\emph{New Key Agreement Protocols in Braid Group Cryptography},
CT-RSA 2001, Lecture Notes in Computer Science \textbf{2020} (2001), 13--27.

\bibitem{BBS09}
L\'aszl\'o Babai, Robert Beals, \'Akos Seress,
\emph{Polynomial-time theory of matrix groups}, ACM STOC 2009, 55--64.

\bibitem{Bigelow01} S. Bigelow, \emph{Braid groups are linear},
Journal of the American Mathematical Society \textbf{14} (2001), 471--486.

\bibitem{BirmanBrendle} J. Birman, T. Brendle, \emph{Braids: A Survey}, in: W. Menasco, M. Thistlethwaite (eds.), 
\textbf{Handbook of Knot Theory}, Elsevier, Amsterdam, 2005, pp.\ 19--103.

\bibitem{CK+01}
J. Cha, K. Ko, S. Lee, J. Han, J. Cheon,
\emph{An efficient implementation of braid groups},
ASIACRYPT 2001, LNCS \textbf{2248} (2001), 144--156.

\bibitem{CJ03}
J. Cheon, B. Jun,
\emph{A polynomial time algorithm for the braid Diffie-Hellman conjugacy problem},
CRYPTO 2003, Lecture Notes in Computer Science \textbf{2729} (2003), 212--224.

\bibitem{DehSurv}
P. Dehornoy,
\emph{Braid-based cryptography},
Contemporary Mathematics \textbf{360} (2004), 5--33.

\bibitem{GarSurv}
D. Garber,
\emph{Braid group cryptography},
in: J. Berrick, F.R. Cohen, E. Hanbury, Y.L. Wong, J. Wu, eds.,
\textbf{Braids: Introductory Lectures on Braids, Configurations and Their Applications},
IMS Lecture Notes Series \textbf{19}, National University of Singapore, 2009, 329--403.

\bibitem{BGG05}
D. Garber, S. Kaplan, M. Teicher, B. Tsaban, U. Vishne,
\emph{Probabilistic solutions of equations in the braid group},
Advances in Applied Mathematics \textbf{35} (2005), 323--334.

\bibitem{Geb05} V. Gebhardt,
\emph{A new approach to the conjugacy problem in Garside groups},
Journal of Algebra \textbf{292} (2005), 282--302.

\bibitem{Geb06} V. Gebhardt,
\emph{Conjugacy search in braid groups},
Applicable Algebra in Engineering, Communication and Computing \textbf{17} (2006), 219--238.

\bibitem{GMMU08} R. Gilman, A. Miasnikov, A. Miasnikov, A. Ushakov,
\emph{New developments in Commutator Key Exchange},
Proceedings of the First International Conference on Symbolic Computation and Cryptography,
Beijing, 2008, 146--150.\\
\texttt{http://www-calfor.lip6.fr/~jcf/Papers/scc08.pdf}

\bibitem{HofSte03}
D. Hofheinz, R. Steinwandt,
\emph{A practical attack on some braid group based cryptographic primitives},
PKC 2003, Lecture Notes in Computer Science \textbf{2567} (2002), 187--198.

\bibitem{HT02} J. Hughes, A. Tannenbaum,
\emph{Length-based attacks for certain group based encryption rewriting systems},
SECI02: S\'ecurit\'e de la Communication sur Internet, 2002.\\
\texttt{www.ima.umn.edu/preprints/apr2000/1696.pdf}

\bibitem{Hughes02} J. Hughes,
\emph{A linear algebraic attack on the AAFG1 braid group cryptosystem},
Information Security And Privacy, Lecture Notes in Computer Science \textbf{2384} (2002), 107--141.

\bibitem{Ka06} A. Kalka,
\emph{Representation attacks on the braid Diffie-Hellman public key encryption},
Applicable Algebra in Engineering, Communication and Computing \textbf{17} (2006), 257--266.

\bibitem{Ka07} A. Kalka,
\emph{Representations of braid groups and braid-based cryptography}, PhD thesis,
Ruhr-Universit\"at Bochum, 2007.\\
\texttt{www-brs.ub.ruhr-uni-bochum.de/netahtml/HSS/Diss/KalkaArkadiusG/}

\bibitem{Ka12} A. Kalka,
\emph{Non-associative public key cryptography}, arXiv eprint 1210.8270, 2012.

\bibitem{KL+00} K. Ko, S. Lee, J. Cheon, J. Han, J. Kang, C. Park,
\emph{New public-key cryptosystem using braid groups},
CRYPTO 2000, Lecture Notes in Computer Science \textbf{1880} (2000),  166--183.

\bibitem{KLT07}
K. Ko, J. Lee, T. Thomas,
\emph{Towards generating secure keys for braid cryptography},
Design Codes and Cryptography \textbf{45} (2007), 317--333.

\bibitem{Krammer02} D. Krammer,
\emph{Braid groups are linear}, Annals of Mathematics \textbf{155} (2002), 131--156.

\bibitem{Kurt} Y. Kurt 
\emph{A new key agreement scheme based on the triple decomposition problem}, 
International Journal of Network Security \textbf{16} (2014), 340--350.

\bibitem{LL02} S. Lee, E. Lee,
\emph{Potential weaknesses of the commutator key agreement protocol based on braid groups},
EUROCRYPT 2002, Lecture Notes in Computer Science \textbf{2332} (2002), 14--28.

\bibitem{Maffre06} S. Maffre,
\emph{A weak key test for braid-based cryptography},
Design Codes and Cryptography \textbf{39} (2006), 347--373.

\bibitem{MSU05} A. Miasnikov, V. Shpilrain, A. Ushakov,
\emph{A practical attack on some braid group based cryptographic protocols},
CRYPTO 2005, Lecture Notes in Computer Science \textbf{3621} (2005), 86--96.

\bibitem{MSU06} A. Miasnikov, V. Shpilrain, A. Ushakov,
\emph{Random subgroups of braid groups: an approach to cryptanalysis of a braid group based cryptographic protocol},
PKC 2006, Lecture Notes in Computer Science \textbf{3958} (2006), 302--314.

\bibitem{MSUbook}
A. Miasnikov, V. Shpilrain, A. Ushakov,
\textbf{Non-commutative Cryptography and Complexity of Group-theoretic Problems},
American Mathematical Society Surveys and Monographs \textbf{177}, 2011.

\bibitem{MU07} A. Miasnikov, A. Ushakov,
\emph{Length based attack and braid groups: cryptanalysis of Anshel-Anshel-Goldfeld key exchange protocol},
PKC 2007, Lecture Notes in Computer Science \textbf{4450} (2007), 76--88.

\bibitem{MU08} A. Myasnikov, A. Ushakov,
\emph{Random subgroups and analysis of the length-based and quotient attacks},
Journal of Mathematical Cryptology \textbf{2} (2008), 29--61.

\bibitem{Lat} D. Micciancio, O. Regev, \emph{Lattice-based Cryptography}, in:
\textbf{Post-quantum Cryptography} (D. Bernstein and J. Buchmann, eds.), Springer, 2008.

\bibitem{LubLec01} A. Lubotzky, \emph{Braid group cryptography}, CS Theory Seminar,
Hebrew University, March 2001.\\
\texttt{http://www.cs.huji.ac.il/~theorys/2001/Alex\_Lubotzky}

\bibitem{Sh08}
V. Shpilrain, \emph{Cryptanalysis of Stickel's key exchange scheme},
in: \textbf{Computer Science in Russia}, Lecture Notes in Computer Science 5010 (2008), 283--288.

\bibitem{ShpUshThomp}
V.\ Shpilrain, A.\ Ushakov,
\emph{Thompson's group and public key cryptography},
ACNS 2005, Lecture Notes in Computer Science \textbf{3531} (2005), 151--164.

\bibitem{ShUsh06}
V.\ Shpilrain, A.\ Ushakov,
\emph{A new key exchange protocol besed on the decomposition problem},
in: L. Gerritzen, D. Goldfeld, M. Kreuzer, G. Rosenberger and V. Shpilrain, eds.,
\textbf{Algebraic Methods in Cryptography},
Contemporary Mathematics \textbf{418} (2006), 161--167.

\bibitem{Stickel}
E. Stickel, 
\emph{A new method for exchanging secret keys}, Proceedings of the Third International Conference on Information Technology and Applications (ICITA05),
2005, 426--430.

\bibitem{GagtaTalk}
B. Tsaban, 
\emph{The Conjugacy Problem: cryptoanalytic approaches to a problem of Dehn}, minicourse, 
D\"usseldorf University, Germany, July--August 2012.\\
\texttt{http://reh.math.uni-duesseldorf.de/\~{}gcgta/slides/Tsaban\_minicourses.pdf}


\end{thebibliography}
